# Entropy and weak solutions in the thermal model for the compressible Euler equations

Zheng Ran and Yupeng Xu

Shanghai Institute of Applied Mathematics and Mechanics, Shanghai University, Shanghai 200072, China

Among the existing models for compressible fluids, the one by Kataoka and Tsutahara (KT model, Phys. Rev. E 69, 056702, 2004) has a simple and rigorous theoretical background. The drawback of this KT model is that it can cause numerical instability if the local Mach number exceeds 1. The precise mechanism of this instability has not yet been clarified. In this paper, we derive entropy functions whose local equilibria are suitable to recover the Euler-like equations in the framework of the lattice Boltzmann method for the KT model. Numerical examples are also given, which are consistent with the above theoretical arguments, and show that the entropy condition is not fully guaranteed in KT model. The negative entropy may be the inherent cause for the non-physical oscillations in the vicinity of the shock. In contrast to these Karlin's microscopic entropy approach, the corresponding subsidiary entropy condition in the LBM calculation could also be deduced explicitly from the macroscopic version, which provides some insights on the numerical instability of the lattice Boltzmann model for shock calculation.

**PACS number(s)**：  47.11. +j, 51.10. +y

## I   INTRODUCTION

It is a challenge to use pure lattice Bolztmann method to simulate the compressible Euler equations, especially for the problem contains shock waves and contact discontinuities [1]. The typical LBM is limited to the low Mach flows, although several related compressible techniques have been proposed to overcome some of these drawbacks [2]. In recent years, a series of LBM for compressible flows has been proposed. Alexander et al. [3] chose a modified equilibrium distribution, allowing the sound speed to be small. Nadiga [4] proposed a discrete velocity model. Huang et al. [5] used flow-adapted discrete velocities, a non-unique equilibrium distribution constrained by a set of linear moments and used interpolated nodes. Prendergast and Xu [6], Kim et al. [7] and Koltelnikov and Montgomery used Bhatnagar-Gross-Krook type model [8] to establish new type flux and employed TVD flux limitation with the neighborhood cells, Rena et al. [9], Vahala et al. [10] , Qian [11,12],Sun [13], De Cicco et al. [14], Mason [15,16], Yan [17,18], and Kataoka and Tsutahara [19,20] proposed many models by using additional techniques to achieve higher Mach number for the compressible flows[21]. Among the existing models for the compressible flows, the one by Kataoka and Tsutahara (KT) has a simple and rigorous theoretical background [22]. It takes flexible ratio of specific heat and is superior in computational efficiency. But similar to previous LB models, the numerical stability problem remains one of the few blocks for its practical simulation to higher Mach number compressible flows. The KT model can cause numerical instability if the local Mach number exceeds 1. The precise mechanism of this instability has not yet been clarified. The reason for this numerical instability is, therefore, a pending problem of the lattice Boltzmann method that should be clarified.

The main idea of this paper could be listed as follows: The spirit of the different LBM approach is to retain the simplest microscopic description that gives the macroscopic behavior of interest. Application of a Taylor series expansion of the lattice kinetic equation followed by a Chapman-Enskog expansion results in the typical hierarchy of equations; Euler, Navier-Stokes, Burnett, etc. By selecting the appropriate number of speeds and the appropriate form of the equilibrium distribution function, one may match the equations that results from the LB method with those of traditional kinetic theory to the desired level. Higher level terms that are not matched represent behavior of the lattice gas that differs from a Maxwellian gas. This brings us to an alternative view that the higher order terms in the Taylor series expansion of the kinetic equation are not "physical" but may be considered "truncation error" of a finite difference approximation to some continuous equation. The governing equations of gas-dynamics are expressions of conservation and the second law of thermodynamics. Conservation requires that three fundamental quantities – mass, momentum, and energy – are neither created nor destroyed but are only redistributed or, excepting mass, converted from one form to another. A companion principle to conservation, known as the second law of thermodynamics, requires that a fourth fundamental quantity called entropy should never decrease. The second law of thermodynamics restricts the redistributions and conversions of conserved quantities otherwise allowed by the conservation laws. As another supplement to conservation, the equations of state, the second law of thermo-dynamics collectively constitute of the Euler equations. An interesting problem we asked: How to realize this counterparts in the thermal LBGK model ? This is the main topic of this paper. It is believed that this is first study that has laid the theoretical foundation of the thermal LBM (KT model) for the simulation of flows with shock waves and contact discontinuities.

We also noted that: In traditional kinetic theory, the equilibrium velocity distribution function is the maximum entropy state. Thus, any initial state will evolve towards a state of higher entropy. This result is known as Boltzmann's H-theorem which ensure an increase of entropy, and ensure stability. An H-theorem has been derived for some particle methods and derivation for lattice gases is included in Ref. [23]. If one can guarantee that the equilibrium distribution function for LBM is the maximum entropy state, then stability can be guaranteed even though LB approaches are not particle methods [24]. The problem with this approach, however, is that one cannot usually find an equilibrium distribution function that simultaneously guarantee an H-theorem and allow the correct form of the equations to be obtained. It is well known that flows with shock waves and contact discontinuities can be correctly described by the weak solutions of the compressible Euler equations with the subsidiary entropy condition. In contrast to these Karlin 's approach, as we will see below, the corresponding subsidiary entropy condition in the LBM calculation could also be deduced explicitly from the macroscopic version [25], which provides some insights on the numerical instability of the KT model for shock calculation.

## II  LATTICE BOLTZMANN MODEL

We introduce a Lattice Boltzmann model in its dimensional form in Section. II A, and then, in its nondimensional form in Section.II B.[19]

### A. Dimensional expressions

First, we write the compressible Euler equation explicitly:

$$\frac{\partial \rho}{\partial t} + \frac{\partial \rho u_\alpha}{\partial x_\alpha} = 0 \tag{1a}$$

$$\frac{\partial \rho u_\alpha}{\partial t} + \frac{\partial \rho u_\alpha u_\beta}{\partial x_\beta} + \frac{\partial p}{\partial x_\alpha} = 0 \tag{1b}$$

$$\frac{\partial \rho \left(bRT + u_\alpha^2\right)}{\partial t} + \frac{\partial \rho u_\alpha \left(bRT + u_\beta^2\right) + 2pu_\alpha}{\partial x_\alpha} = 0 \tag{1c}$$

$$(\alpha = 1, 2, ..., D; \beta = 1, 2, ..., D),$$

Where $t$ is the time, $x_\alpha$ is the spatial coordinate, $\rho, u_\alpha, T$ and

$$p = \rho RT \tag{2}$$

are, respectively, the density, the flow velocity in the $x_\alpha$ direction, the temperature, and the pressure of a gas. $R$ and $D$ are the specific gas constant and the number of spatial dimensions, respectively. $b$ is a given constant expressed as

$$b = \frac{2}{\gamma - 1} \tag{3}$$

Where $\gamma$ is the specific-heat ratio. Note that, in the present study, the subscripts $\alpha$

and $\beta$ represent the number of spatial coordinates and the summation convection is applied to these subscripts. The initial condition is

$$\rho = \rho^0, \quad u_\alpha = u_\alpha^0 \quad T = T^0 \quad \text{at} \quad t = 0 \tag{4}$$

Where $\rho^0, u_\alpha^0$ and $T^0$ are given functions of $x_\alpha$

Now we present a Lattice Boltzmann model that gives the solutions of the initial-value problem of the compressible Euler equations (1a)-(1c) with the initial condition (4). Let $c_{i\alpha}$ ($i = 0,1,2,...,I-1$; $I$ is the total number of discrete molecular velocities) be the molecular velocity in the $x_\alpha$ direction of the $i$th particle, and $\eta_i$ be another variable newly introduced to control the specific-heat ratio. $f_i(t, x_\alpha)$ is the velocity distribution function of the $i$th particle. The macroscopic variables $\rho$, $u_\alpha$ and $T$ are defined as

$$\rho = \sum_{i=0}^{I-1} f_i \tag{5a}$$

$$\rho u_\alpha = \sum_{i=0}^{I-1} f_i c_{i\alpha} \tag{5b}$$

$$\rho\left(bRT + u_\alpha^2\right) = \sum_{i=0}^{I-1} f_i \left(c_{i\alpha}^2 + \eta_i^2\right) \tag{5c}$$

Note that, in the present study, the summation convection is not applied to the subscript $i$ representing the kind of the molecules

Consider the initial-value problem of the Bhatnger-Gross-Krook-type kinetic equation [26]

$$\frac{\partial f_i}{\partial t} + c_{i\beta} \frac{\partial f_i}{\partial x_\beta} = \frac{f_i^{eq}(\rho, u_\alpha, T) - f_i}{\tau} \tag{6}$$

With the initial condition

$$f_i = f_i^{eq}\left(\rho^0, u_\alpha^0, T^0\right) \quad \text{at} \quad t=0 \tag{7}$$

Where $\tau$ (the relaxation time) is a given constant and $f_i^{eq}(\rho, u, T)$ (The local equilibrium velocity distribution function) is a given function of the macroscopic variables. In the LBM, the following discretized form of Eq.(6) is often used:

$$\frac{f_i(t+\Delta t, x_\alpha + c_{i\alpha}\Delta t) - f_i(t, x_\alpha)}{\Delta t} = \frac{f_i^{eq}(\rho, u_\alpha, T) - f_i}{\tau} \tag{8}$$

Where $\Delta t$ is the discrete time step of order $\tau$. It is clear that Eq.(8) is only one of the finite-difference scheme of the kinetic equation(6). Therefore, we use Eq.(6) as a basic kinetic equation in the following. It is also noted that there is a recent trend in the LBM community to use the usual finite-difference scheme of Eq.(6) rather than the numerical stability problem[27,28,29]

Now return to the explanation of the Lattice Boltzmann models. The following constrains are imposed on the moments of $f_i^{eq}$ appearing on the right-hand sides of Eqs. (6) and (7):

$$\rho = \sum_{i=0}^{I-1} f_i^{eq} \tag{9a}$$

$$\rho u_\alpha = \sum_{i=0}^{I-1} f_i^{eq} c_{i\alpha} \tag{9b}$$

$$p\delta_{\alpha\beta} + \rho u_\alpha u_\beta = \sum_{i=0}^{I-1} f_i^{eq} c_{i\alpha} c_{i\beta} \tag{9c}$$

$$\rho\left(bRT + u_\alpha^2\right) = \sum_{i=0}^{I-1} f_i^{eq}\left(c_{i\alpha}^2 + \eta_i^2\right) \tag{9d}$$

$$\rho\left[(b+2)RT + u_\beta^2\right]u_\alpha = \sum_{i=0}^{I-1} f_i^{eq}\left(c_{i\beta}^2 + \eta_i^2\right)c_{i\alpha} \tag{9e}$$

Then the macroscopic variables $\rho$, $u_\alpha$, and $T$ derived from the solution of the kinetic equation (6) with the initial condition (7) satisfy the compressible Euler equations (1a)-(1c) and their initial condition (4) if the time and length scales of variation of solution are much larger than $\tau$ and $\tau\sqrt{RT}$ ,respectively. The proof is given in Sction.IIIA.

The specific models that satisfy the above constraints (9a)-(9e) are presented in the following section (Sec.IIIB).

### B. Nondimensional expressions

The nondimensional variables and equations, which are convenient for the following analysis and numerical calculation, are listed first.

Let $L$, $\rho_0$ and $T_0$ be, respectively, the reference length, density, and the temperature. Then, the nondimensional variables are defined as follows:

$$\hat{t} = \frac{t}{L/\sqrt{RT_0}}, \quad \hat{x}_\alpha = \frac{x_\alpha}{L}, \quad \hat{c}_{i\alpha} = \frac{c_{i\alpha}}{\sqrt{RT_0}}, \quad \hat{\eta}_i = \frac{\eta_i}{\sqrt{RT_0}}$$

$$\hat{f}_i = \frac{f_i}{\rho_0}, \quad \hat{f}_i^{eq} = \frac{f_i^{eq}}{\rho_0},$$

$$\hat{\rho} = \frac{\rho}{\rho_0}, \quad \hat{u}_\alpha = \frac{u_\alpha}{\sqrt{RT_0}}, \hat{T} = \frac{T}{T_0}, \hat{p} = \frac{p}{\rho_0 RT_0},$$

$$\hat{\rho}^0 = \frac{\rho^0}{\rho_0}, \quad \hat{u}_\alpha^0 = \frac{u_\alpha^0}{\sqrt{RT_0}}, \hat{T}^0 = \frac{T^0}{T_0}, \hat{p}^0 = \frac{p^0}{\rho_0 RT_0}. \tag{10}$$

In terms of these nondimensional variables, the compressible Euler equations (1a)-(1c) and their initial condition (4) are

$$\frac{\partial \hat{\rho}}{\partial \hat{t}} + \frac{\partial \hat{\rho}\hat{u}_\alpha}{\partial \hat{x}_\alpha} = 0 \tag{11a}$$

$$\frac{\partial \hat{\rho}\hat{u}_\alpha}{\partial \hat{t}} + \frac{\partial \hat{\rho}\hat{u}_\alpha \hat{u}_\beta}{\partial \hat{x}_\beta} + \frac{\partial \hat{p}}{\partial \hat{x}_\alpha} = 0 \tag{11b}$$

$$\frac{\partial \hat{\rho}\left(b\hat{T}+\hat{u}_{\alpha}^{2}\right)}{\partial \hat{t}}+\frac{\partial \hat{\rho}\hat{u}_{\alpha}\left(b\hat{T}+\hat{u}_{\beta}^{2}\right)+2\hat{p}\hat{u}_{\alpha}}{\partial \hat{x}_{\alpha}}=0 \tag{11c}$$

$$(\alpha=1,2,...,D;\beta=1,2,...,D),$$

Where

$$\hat{p}=\hat{\rho}\hat{T} \tag{12}$$

and

$$\hat{\rho}=\hat{\rho}^{0},\ \hat{u}_{\alpha}=\hat{u}_{\alpha}^{0}\quad \hat{T}=\hat{T}^{0}\quad \text{at}\quad \hat{t}=0 \tag{13}$$

The nondimensional macroscopic variables used in the LBM are defined as

$$\hat{\rho}=\sum_{i=0}^{I-1}\hat{f}_{i} \tag{14a}$$

$$\hat{\rho}\hat{u}_{\alpha}=\sum_{i=0}^{I-1}\hat{f}_{i}\hat{c}_{i\alpha} \tag{14b}$$

$$\hat{\rho}\left(b\hat{T}+\hat{u}_{\alpha}^{2}\right)=\sum_{i=0}^{I-1}\hat{f}_{i}\left(\hat{c}_{i\alpha}^{2}+\hat{\eta}_{i}^{2}\right) \tag{14c}$$

From Eqs.(5a)-(5c).the kinetic equation (6) and its initial condition (7) of nondimensional form are

$$\frac{\partial \hat{f}_{i}}{\partial t}+\hat{c}_{i\beta}\frac{\partial \hat{f}_{i}}{\partial \hat{x}_{\beta}}=\frac{\hat{f}_{i}^{eq}\left(\rho,u_{\alpha},T\right)-\hat{f}_{i}}{\varepsilon} \tag{15}$$

And

$$\hat{f}_{i}=\hat{f}_{i}^{eq}\left(\hat{\rho}^{0},\hat{u}_{\alpha}^{0},\hat{T}^{0}\right)\quad \text{at}\quad \hat{t}=0 \tag{16}$$

Where $\varepsilon$ is the Knudsen number defined by

$$\varepsilon=\frac{\tau\sqrt{RT_{0}}}{L} \tag{17}$$

$\hat{f}_{i}^{eq}$ satisfies the following constraints from Eqs.(9a)-(9e):

$$\hat{\rho}=\sum_{i=0}^{I-1}\hat{f}_{i}^{eq} \tag{18a}$$

$$\hat{\rho}\hat{u}_\alpha = \sum_{i=0}^{I-1} \hat{f}_i^{eq} \hat{c}_{i\alpha} \tag{18b}$$

$$\hat{p}\delta_{\alpha\beta} + \hat{\rho}\hat{u}_\alpha \hat{u}_\beta = \sum_{i=0}^{I-1} \hat{f}_i^{eq} \hat{c}_{i\alpha} \hat{c}_{i\beta} \tag{18c}$$

$$\hat{\rho}\left(b\hat{T} + \hat{u}_\alpha^2\right) = \sum_{i=0}^{I-1} \hat{f}_i^{eq} \left(\hat{c}_{i\alpha}^2 + \hat{\eta}_i^2\right) \tag{18d}$$

$$\hat{\rho}\left[(b+2)\hat{T} + \hat{u}_\beta^2\right]\hat{u}_\alpha = \sum_{i=0}^{I-1} \hat{f}_i^{eq} \left(\hat{c}_{i\beta}^2 + \hat{\eta}_i^2\right)\hat{c}_{i\alpha} \tag{18e}$$

We will give a one-dimensional model that satisfies the above constrains (18a)-(18e)

### III. ONE DIMENSIONAL MODEL AND ITS EULER EQUATIONS

The lattice Boltzmann model for the compressible Euler equations has been proposed together with its rigorous theoretical background by Kataoka and Tsutahara in 2004. Compared with the previous method in the thermal LBM, the KT model has completely overcome the defects that the specific heat ratio cannot be chosen freely. In order to make a discussion simple, we only consider the one-dimensional problem in this paper. Following the KT model, for the one-dimensional model (D=1, I=5), we have

$$\hat{c}_{i1} = \begin{cases} 0 & i=0 \\ v_1 \cos\left[(i-1)\pi\right] & i=1,2 \\ v_2 \cos\left[(i-1)\pi\right] & i=3,4 \end{cases}$$

(19)

$$\hat{\eta}_i = \begin{cases} \eta_0 & i=0 \\ 0 & i=1,2,3,4 \end{cases}$$

Where $v_1$, $v_2 (\neq v_1)$, and $\eta_0$ are given nonzero constants, and let

$$\hat{f}_i^{eq} = \hat{\rho}\left(A_i + B_i \hat{u}_1 \hat{c}_{i1}\right) \qquad \text{for } i = 0,1,2,3,4 \tag{20}$$

be a local equilibrium velocity distribution function, where

$$A_i = \begin{cases} \dfrac{b-1}{\eta_0^2}\hat{T} & i=0 \\[2mm] \dfrac{1}{2(v_1^2 - v_2^2)}\left[-v_2^2 + \left((b-1)\dfrac{v_2^2}{\eta_0^2} + 1\right)\hat{T} + \hat{u}_1^2\right] & i=1,2 \\[2mm] \dfrac{1}{2(v_2^2 - v_1^2)}\left[-v_1^2 + \left((b-1)\dfrac{v_1^2}{\eta_0^2} + 1\right)\hat{T} + \hat{u}_1^2\right] & i=3,4 \end{cases} \tag{21a}$$

$$B_i = \begin{cases} \dfrac{-v_2^2 + (b+2)\hat{T} + \hat{u}_1^2}{2v_1^2(v_1^2 - v_2^2)} & i=1,2 \\[2mm] \dfrac{-v_1^2 + (b+2)\hat{T} + \hat{u}_1^2}{2v_2^2(v_2^2 - v_1^2)} & i=3,4 \end{cases} \tag{21b}$$

Then $\hat{c}_{i1}$, $\hat{\eta}_i$ and $\hat{f}_i^{eq}$ $(i=0,1,2,3,4)$ given above satisfy the constrains (18a)-(18e).This is the first lattice Boltzmann model of one-dimensional version whose specific-heat ratio $\gamma$ [which is related to $b$ by Eq. (3)] can be chosen according to our convenience, while the previous model gives the unphysical value of $\gamma = 3$ (or $b=1$) only

Through the Chapman-Enskog analysis (APPENDIX),we obtain

$$\partial_{t_0} f_i^{eq} + c_{i\beta}\partial_\beta f_i^{eq} = -\omega f_i^{(1)} \tag{22a}$$

$$\partial_{t_0} f_i^{(1)} + \partial_{t_1} f_i^{eq} + c_{i\beta}\partial_\beta f_i^{(1)} + \frac{1}{2}\partial_{t_0 t_0} f_i^{eq} + \frac{1}{2}c_{i\beta}c_{i\gamma}\partial_{\beta\gamma} f_i^{eq} + c_{i\beta}\partial_{t_0\beta} f_i^{eq} = -\omega f_i^{(2)} \tag{22b}$$

The leading order terms are the Euler equations of perfect gas with the truncation errors $R_i = O(\varepsilon)$. (APPENDIX)

$$\partial_t \rho + \partial_x (\rho u) = R_1 \tag{23a}$$

$$\partial_t(\rho u) + \partial_x(\rho u^2 + P) = R_2 \tag{23b}$$

$$\partial_t(\rho u^2 + \rho bT) + \partial_x \rho\left[(b+2)T + u^2\right]u = R_3 \tag{23c}$$

Where

$$R_1 = 0 \tag{24a}$$

$$R_2 = \frac{\delta}{\omega}\left[\frac{2(b^2-1)}{b}\frac{\partial p}{\partial x}\frac{\partial u}{\partial x} + (b-1)u\frac{\partial^2 p}{\partial x^2} + \frac{(b+2)(b-1)}{b}p\frac{\partial^2 u}{\partial x^2}\right] \tag{24b}$$

$$R_3 = \frac{\delta}{\omega}\left\{\begin{array}{l} -\left[\dfrac{2(b+1)(b+2)}{b}p + 12\rho u^2 + 2\rho(v_1^2 + v_2^2)\right]\left(\dfrac{\partial u}{\partial x}\right)^2 \\[6pt] -\dfrac{4(b^2+4b+1)}{b}u\dfrac{\partial p}{\partial x}\dfrac{\partial u}{\partial x} - \left[8u^3 - 4u(v_1^2 + v_2^2)\right]\dfrac{\partial \rho}{\partial x}\dfrac{\partial u}{\partial x} \\[6pt] +(b+2)\dfrac{p}{\rho^2}\dfrac{\partial \rho}{\partial x}\dfrac{\partial p}{\partial x} - (b+2)\dfrac{1}{\rho}\dfrac{\partial p}{\partial x}\dfrac{\partial p}{\partial x} \\[6pt] -\left[\dfrac{2(b+2)(b+1)}{b}pu + 4\rho u^3 + 2\rho u(v_1^2 + v_2^2)\right]\dfrac{\partial^2 u}{\partial x^2} \\[6pt] -\left[(b+5)u^2 + (b+2)\dfrac{p}{\rho} + \dfrac{(b-1)v_1^2 v_2^2}{\eta_0^2} + (v_1^2 + v_2^2)\right]\dfrac{\partial^2 p}{\partial x^2} \\[6pt] -\left[u^4 + v_1^2 v_2^2 + (v_1^2 + v_2^2)u^2\right]\dfrac{\partial^2 \rho}{\partial x^2} \end{array}\right. \tag{24c}$$

## IV. THE DERIVATION OF THE ENTROPY EQUATION

In Ref. [24], the derivation of the entropy condition for LBGK model has been presented. By using the same manipulation, one could obtain the corresponding result for KT model. The derivation of the entropy equation from the Eqs.(23a)-(23c) in the one-dimensional problem as follows. Rewrite Eqs. (23a)- (23c) as

$$\partial_t \rho + \partial_x(\rho u) = 0 \tag{25a}$$

$$\partial_t(\rho u) + \partial_x(\rho u^2 + P) = R_2 \tag{25b}$$

$$\partial_t(\rho E) + \partial_x(pu + \rho Eu) = R_3 \tag{25c}$$

where

$$E = e + \frac{1}{2}u^2 \tag{26}$$

Eq. (25c)- $E \times$ Eq. (25a), we have

$$\rho \partial_t E + \rho u \partial_x E + \partial_x(Pu) = R_3 \tag{27}$$

Then substituting Eq. (26) into Eq. (27), we have

$$\partial_t e + u \partial_x e = \frac{1}{\rho}\left[R_3 - \rho \partial_t\left(\frac{1}{2}u^2\right) - \rho u \partial_x\left(\frac{1}{2}u^2\right) - \partial_x(Pu)\right] \tag{28}$$

Based on the definition of entropy

$$dS = dQ/T \tag{29}$$

It is well known that

$$de = dQ + \frac{P}{\rho^2}d\rho \tag{30}$$

Hence

$$TdS = de - \frac{P}{\rho^2}d\rho \tag{31}$$

At this stage, we have

$$\partial_t S + u \partial_x S = \frac{1}{T}\left[(\partial_t e + u \partial_x e) - \frac{P}{\rho^2}(\partial_t \rho + u \partial_x \rho)\right] \tag{32}$$

The substituting Eq. (25a) into Eq. (32), we obtain

$$\partial_t S + u \partial_x S = \frac{1}{T}\left(\partial_t e + u \partial_x e + \frac{P}{\rho}\partial_x u\right) \tag{33}$$

At last we could deduce the entropy equation as following

$$\partial_t S + u \partial_x S = \frac{1}{\rho T}\left[R_3 - \rho \partial_t\left(\frac{1}{2}u^2\right) - \rho u \partial_x\left(\frac{1}{2}u^2\right) - \partial_x(Pu) + P \partial_x u\right]$$

$$= \frac{1}{\rho T}\left[R_3 - \rho u \partial_t u - \rho u^2 \partial_x u - u \partial_x P\right]$$

$$= \frac{1}{\rho T}\left[R_3 - u(\rho \partial_t u) - \rho u^2 \partial_x u - u \partial_x P\right]$$

$$= \frac{1}{\rho T}\left\{R_3 - u\left[R_2 - \partial_x P - \partial_x(\rho u^2) - u \partial_t \rho\right] - \rho u^2 \partial_x u - u \partial_x P\right\}$$

$$= \frac{1}{\rho T}\left\{R_3 - u\left[R_2 - \partial_x P - \partial_x(\rho u^2) + u \partial_x(\rho u)\right] - \rho u^2 \partial_x u - u \partial_x P\right\}$$

$$= \frac{1}{\rho T}\left\{R_3 - u R_2 + u \partial_x P + u \partial_x(\rho u^2) - u^2 \partial_x(\rho u) - \rho u^2 \partial_x u - u \partial_x P\right\}$$

$$= \frac{1}{\rho T}\left\{R_3 - u R_2 + u \partial_x(\rho u^2) - u^2 \partial_x(\rho u) - \rho u^2 \partial_x u\right\}$$

$$= \frac{1}{\rho T}\left\{R_3 - u R_2 + u^2 \partial_x(\rho u) + \rho u^2 \partial_x u - u^2 \partial_x(\rho u) - \rho u^2 \partial_x u\right\}$$

$$= \frac{1}{\rho T}(R_3 - u R_2)$$

$$= \frac{\delta}{\omega}\frac{1}{\rho T}\left\{\begin{array}{l}-\left[\dfrac{2(b+1)(b+2)}{b}p + 12\rho u^2 - 2\rho(v_1^2 + v_2^2)\right]\left(\dfrac{\partial u}{\partial x}\right)^2 - \dfrac{2(2b^2 + 6b + 1)}{b} u \dfrac{\partial p}{\partial x}\dfrac{\partial u}{\partial x} \\ + \left[4u(v_1^2 + v_2^2) - 8u^3\right]\dfrac{\partial \rho}{\partial x}\dfrac{\partial u}{\partial x} + (b+2)\dfrac{p}{\rho^2}\dfrac{\partial \rho}{\partial x}\dfrac{\partial p}{\partial x} - (b+2)\dfrac{1}{\rho}\dfrac{\partial p}{\partial x}\dfrac{\partial p}{\partial x} \\ -\left[\dfrac{(b+2)(3b+1)}{b}pu + 4\rho u^3 - 2\rho u(v_1^2 + v_2^2)\right]\dfrac{\partial^2 u}{\partial x^2} \\ -\left[\dfrac{b+2}{b}u^2 + (b+2)\dfrac{p}{\rho} - \dfrac{(b-1)v_1^2 v_2^2}{\eta_0^2} - (v_1^2 + v_2^2)\right]\dfrac{\partial^2 p}{\partial x^2} \\ -\left[u^4 + v_1^2 v_2^2 - u^2(v_1^2 + v_2^2)\right]\dfrac{\partial^2 \rho}{\partial x^2}\end{array}\right\}$$

Now we compute

$$C = \left\{ \begin{array}{l} -\left[ \dfrac{2(b+1)(b+2)}{b} p + 12\rho u^2 - 2\rho\left(v_1^2 + v_2^2\right) \right]\left(\dfrac{\partial u}{\partial x}\right)^2 - \dfrac{2\left(2b^2 + 6b + 1\right)}{b} u \dfrac{\partial p}{\partial x}\dfrac{\partial u}{\partial x} \\ + \left[ 4u\left(v_1^2 + v_2^2\right) - 8u^3 \right]\dfrac{\partial \rho}{\partial x}\dfrac{\partial u}{\partial x} + (b+2)\dfrac{p}{\rho^2}\dfrac{\partial \rho}{\partial x}\dfrac{\partial p}{\partial x} - (b+2)\dfrac{1}{\rho}\dfrac{\partial p}{\partial x}\dfrac{\partial p}{\partial x} \\ -\left[ \dfrac{(b+2)(3b+1)}{b} pu + 4\rho u^3 - 2\rho u\left(v_1^2 + v_2^2\right) \right]\dfrac{\partial^2 u}{\partial x^2} \\ -\left[ \dfrac{b+2}{b} u^2 + (b+2)\dfrac{p}{\rho} - \dfrac{(b-1)v_1^2 v_2^2}{\eta_0^2} - \left(v_1^2 + v_2^2\right) \right]\dfrac{\partial^2 p}{\partial x^2} \\ -\left[ u^4 + v_1^2 v_2^2 - u^2\left(v_1^2 + v_2^2\right) \right]\dfrac{\partial^2 \rho}{\partial x^2} \end{array} \right\}$$

to demonstrate the entropy preservation condition of the model.

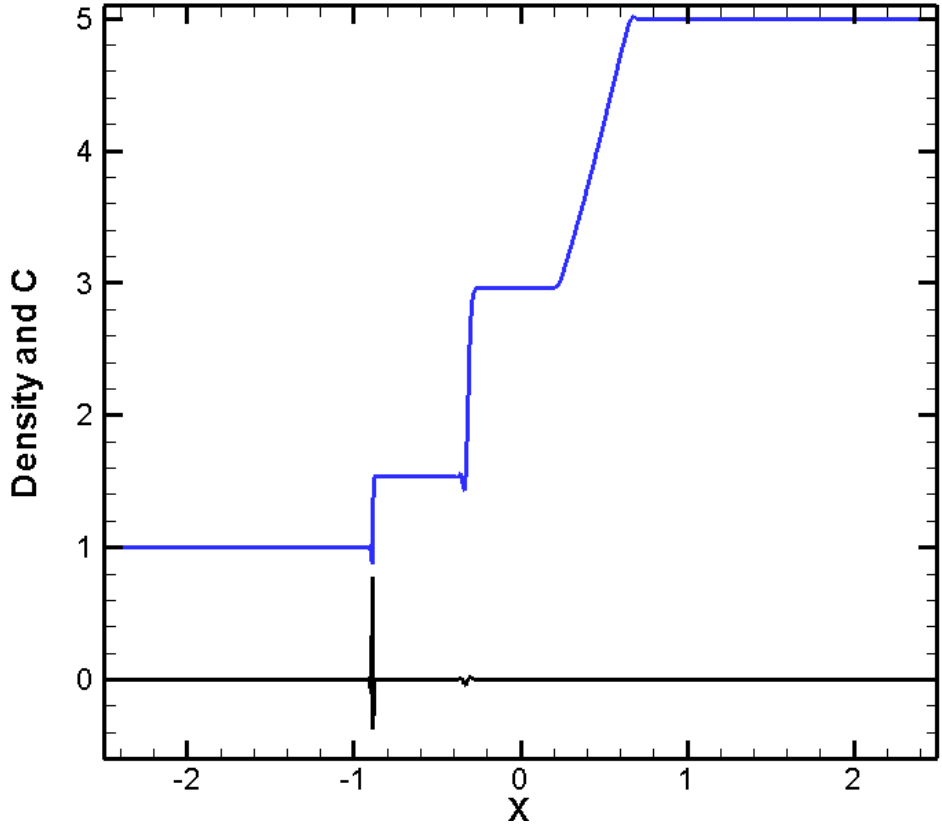

**Figure 1** the density and entropy profiles with $\gamma = 5/3$, $t = 0.5$

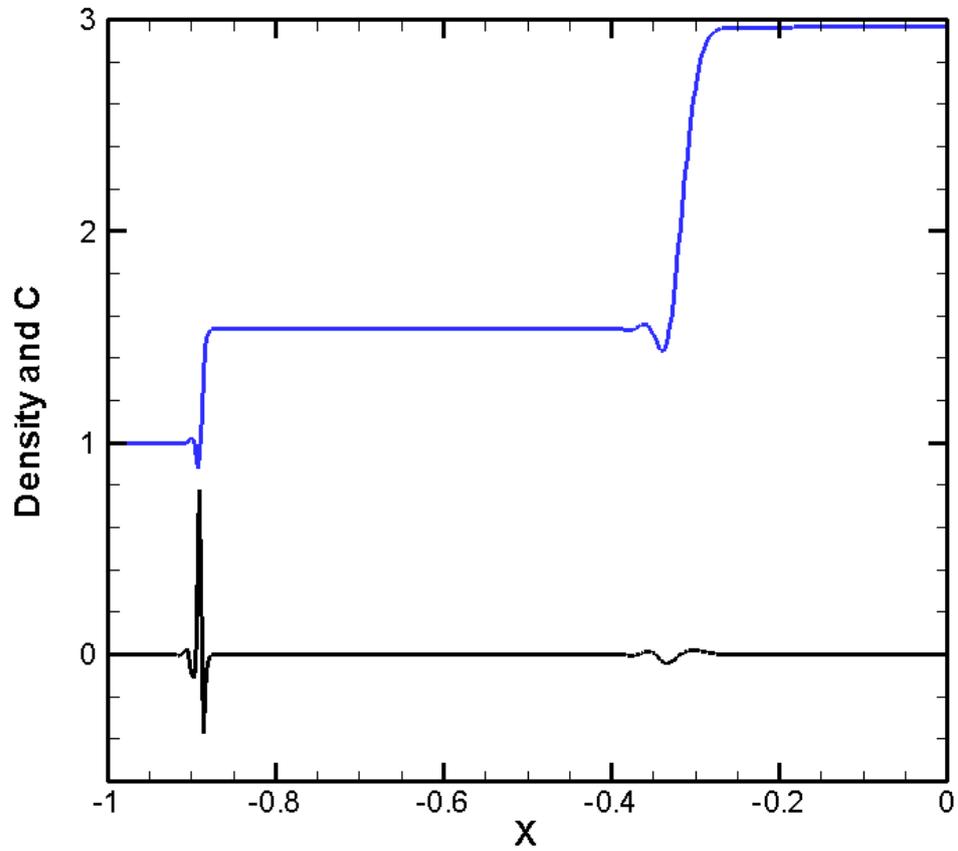

**Figure 2** the density and entropy profiles(local amplification) with $\gamma = 5/3$, $t = 0.5$

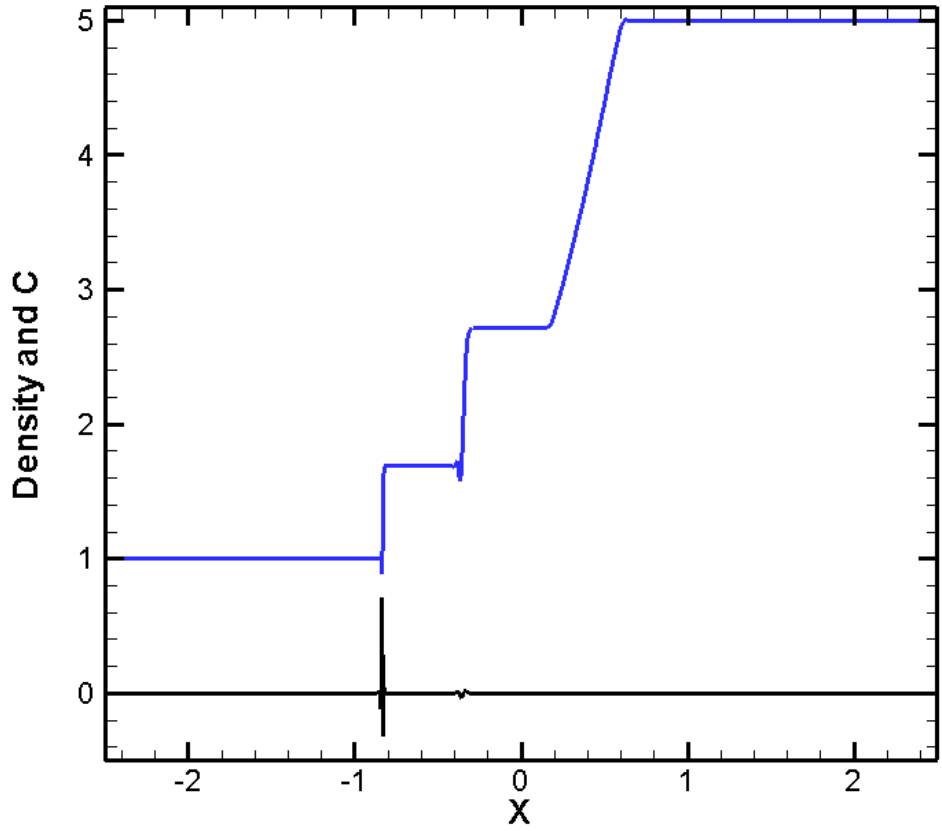

Figure3 the density and entropy profiles with $\gamma = 7/5$, $t = 0.5$

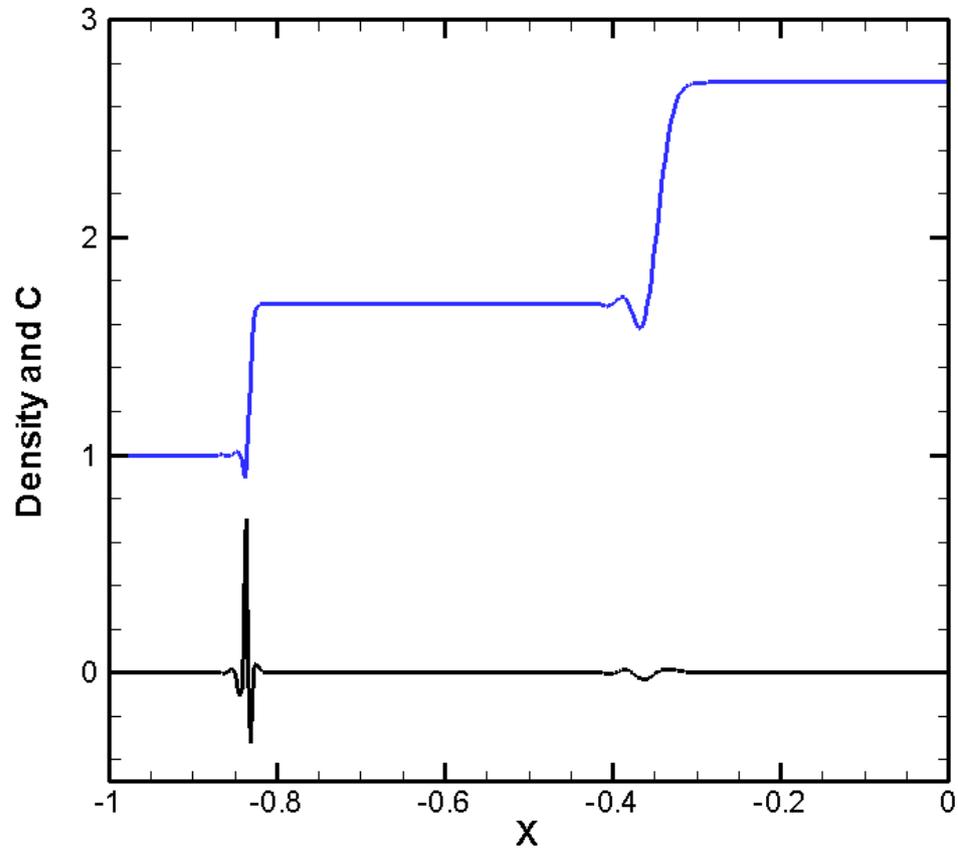

Figure 4 the density and entropy profiles(local amplification) with $\gamma = 7/5$, $t = 0.5$

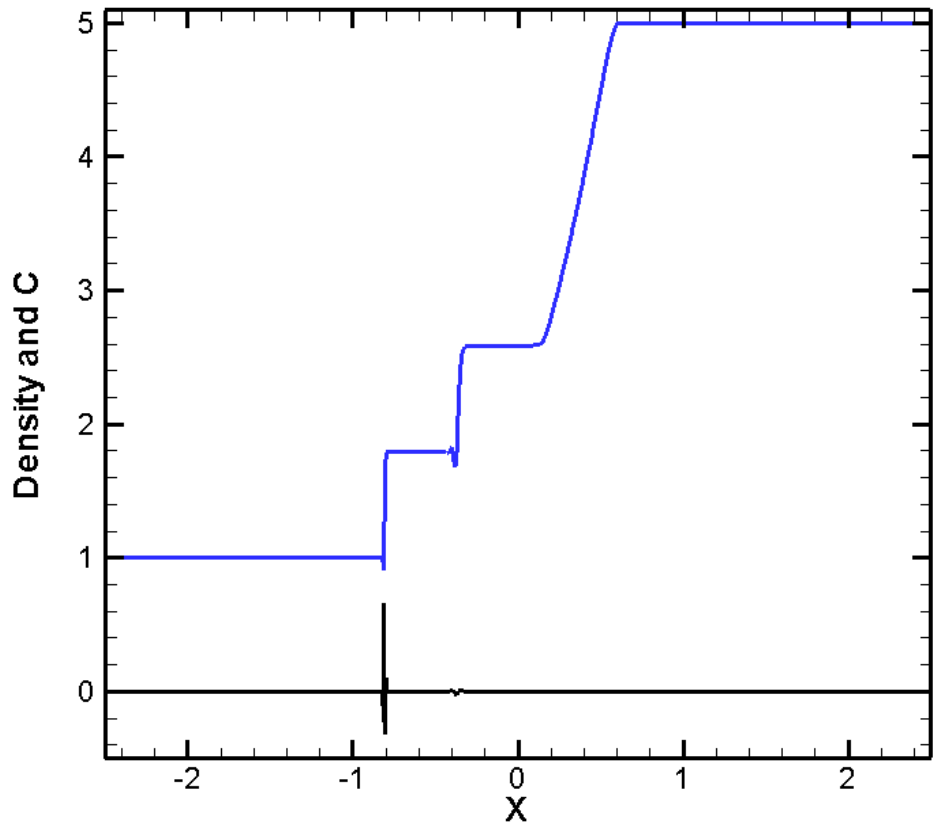

**Figure5 the density and entropy profiles with** $\gamma = 9/7$, $t = 0.5$

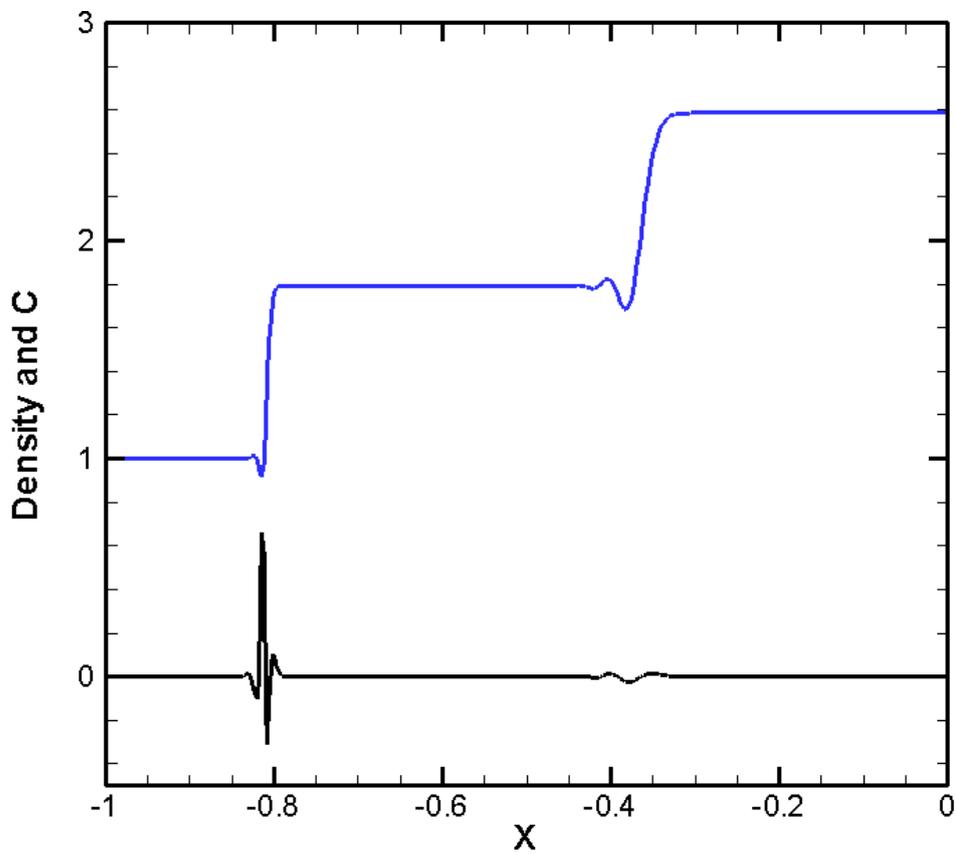

**Figure 6 the density and entropy profiles(local amplification) with** $\gamma = 9/7$, $t = 0.5$

## V. DISCUSSION AND CONCLUSIONS

Now we make some remarks on the numerical results of the KT model together with the corresponding entropy calculation. In this paper, we focus our attention on the Riemann problem. We use the one-dimensional five velocities model (D1Q5) to numerically study the entropy in the KT model [19] calculating. From Fig.1 to Fig.6, the entropy profiles are presented at different and different $\gamma$ ranging from $5/3$ to $9/7$. From the above numerical examples, we find that : In all cases, we observe a negative entropy range exist near the shock, while oscillatory are captured. This indicates that the entropy condition is not fully guaranteed. Thus, the shock waves across which the entropy of fluid particle decreases may appear in the KT model results. As a result, it is not surprising that the numerical instability have been frequently encountered by using the KT model. We reported in this paper the entropy calculation in the KT model. Great care has been paid to the links between the oscillation solution and the entropy negative. An explicit relation is obtained and confirmed by the numerical simulation, which must be in accordance with the second law of thermodynamics. The entropy condition for general Lattice Boltzmann method is quite interesting and debatable question. How to use this find to improve the thermal LBM is our further study.

## APPENDIX: CHAPMAN-ENSKOG PROCEDURE AND RECOVERY OF EULER EQUATION

First, we write the evolution equation of LBE and substituting Knudsen number and time step with $\delta$, then

$$f_i(\mathbf{x}+\delta\mathbf{e}_i,t+\delta)-f_i(\mathbf{x},t)=\Omega_i \tag{A1}$$

Choose the BGK collision term

$$\Omega_i = \omega\left(f_i^{eq}-f_i\right) \tag{A2}$$

Where $i=1,2\cdots b$, $f_i^{eq}$ is the equilibrium distribution function, then carry the Taylor expansion on the left hand of Eq. (A1)

$$f_i(\mathbf{x}+\delta\mathbf{e}_i,t+\delta)-f_i(\mathbf{x},t)=\sum_{n=1}^{\infty}\frac{\delta^n}{n!}\left(\partial_t+c_{i\alpha}\partial_\alpha\right)^n f_i(\mathbf{x},t) \tag{A3}$$

Up to the second order in expansion parameter $\delta$, the right hand of (A3) becomes

$$\delta\left(\partial_t+c_{i\beta}\partial_\beta\right)f_i+\frac{1}{2}\delta^2\left(\partial_t+c_{i\beta}\partial_\beta\right)^2 f_i+O(\delta^3)=\Omega_i \tag{A4}$$

Now we introduce two time scale $t_0=t$, $t_1=\delta t$ and then use the multiscale technique and Chapman-Enskog expansion

$$f_i=\sum_{n=0}^{\infty}\delta^n f_i^{(n)}=f_i^{eq}+\delta f_i^{(1)}+\delta^2 f_i^{(2)}+O(\delta^3) \tag{A5}$$

$$\partial_t=\partial_{t_0}+\delta\partial_{t_1}+O(\delta^2) \tag{A6}$$

Substituting (A5) and (A6) into (A4), then comparing the same-order coefficient of $\delta$, $\delta^2$, we obtain

$$\partial_{t_0}f_i^{eq}+c_{i\beta}\partial_\beta f_i^{eq}=-\omega f_i^{(1)} \tag{A7}$$

$$\partial_{t_0}f_i^{(1)}+\partial_{t_1}f_i^{eq}+c_{i\beta}\partial_\beta f_i^{(1)}+\frac{1}{2}\partial_{t_0 t_0}f_i^{eq}+\frac{1}{2}c_{i\beta}c_{i\gamma}\partial_{\beta\gamma}f_i^{eq}+c_{i\beta}\partial_{t_0\beta}f_i^{eq}$$
$$=-\omega f_i^{(2)} \tag{A8}$$

Density $\rho$, velocity $u_\beta$ and temperature are defined as $T$

$$\rho=\sum_i f_i=\sum_i f_i^{eq} \tag{A9}$$

$$\rho u_\beta=\sum_i f_i c_{i\beta}=\sum_i f_i^{eq}c_{i\beta} \tag{A10}$$

$$\rho\left(bT+u_\beta^2\right)=\sum_i f_i\left(c_{i\beta}^2+\eta_i^2\right)=\sum_i f_i^{eq}\left(c_{i\beta}^2+\eta_i^2\right) \tag{A11}$$

The collision term satisfy the following constrains:

$$\sum_i \Omega_i = 0 \quad \sum_i c_{i\beta}\Omega_i = 0 \quad \sum_i \left(c_{i\beta}^2+\eta_i^2\right)\Omega_i = 0 \tag{A12}$$

Stressor tensor and the current vector of energy are defined as

$$\Pi_{\alpha\beta}=\sum_i c_{i\alpha}c_{i\beta}f_i \qquad Q_\alpha=\sum_i\left(c_{i\beta}^2+\eta_i^2\right)c_{i\alpha}f_i$$

Introduce the following notation

$$\Pi_{\alpha\beta}^{(0)}=\sum_i c_{i\alpha}c_{i\beta}f_i^{eq} \tag{A13}$$

$$\Pi_{\alpha\beta}^{(1)}=\sum_i c_{i\alpha}c_{i\beta}f_i^{(1)} \tag{A14}$$

$$Q_\alpha^{(0)}=\sum_i\left(c_{i\beta}^2+\eta_i^2\right)c_{i\alpha}f_i^{eq} \tag{A15}$$

$$Q_\alpha^{(1)}=\sum_i\left(c_{i\beta}^2+\eta_i^2\right)c_{i\alpha}f_i^{(1)} \tag{A16}$$

And

$$P_{\alpha\beta\gamma}^{(0)}=\sum_i c_{i\alpha}c_{i\beta}c_{i\gamma}f_i^{eq} \tag{A17}$$

$$R_{\alpha\beta}^{(0)}=\sum_i\left(c_{i\gamma}^2+\eta_i^2\right)c_{i\alpha}c_{i\beta}f_i^{eq} \tag{A18}$$

Multiplying Eq. (7) with $1, c_{i\alpha}, \left(c_{i\alpha}^2+\eta_i^2\right)$, we can obtain

$$\partial_{t_0}\rho+\partial_\beta\left(\rho u_\beta\right)=0 \tag{A19}$$

$$\partial_{t_0}\left(\rho u_\alpha\right)+\partial_\beta\Pi_{\alpha\beta}=0 \tag{A20}$$

$$\partial_{t_0}\rho\left(bT+u_\alpha^2\right)+\partial_\beta Q_\beta^{(0)}=0 \tag{A21}$$

Multiplying Eq. (7) with $c_{i\alpha}c_{i\gamma} \quad \left(c_{i\gamma}^2+\eta_i^2\right)c_{i\alpha}$, we can obtain

$$\Pi_{\alpha\gamma}^{(1)}=-\frac{1}{\omega}\left(\partial_{t_0}\Pi_{\alpha\gamma}^{(0)}+\partial_\beta P_{\beta\alpha\gamma}^{(0)}\right) \tag{A22}$$

$$Q_\alpha^{(1)} = -\frac{1}{\omega}\left(\partial_{t_0} Q_\alpha^{(0)} + \partial_\beta R_{\beta\alpha}^{(0)}\right) \quad (A23)$$

Multiplying Eq. (8) with $1, c_{i\alpha}, (c_{i\alpha}^2 + \eta_i^2)$, then

$$\partial_{t_1}\rho = -\partial_{t_0}\sum_i f_i^{(1)} - \partial_\beta \sum_i c_{i\beta} f_i^{(1)} - \frac{1}{2}\partial_{t_0 t_0}\rho - \frac{1}{2}\partial_{\beta\gamma}\Pi_{\beta\gamma}^{(0)} - \partial_{t_0\beta}(\rho u_\beta) \quad (A24)$$

$$\partial_{t_1}(\rho u_\alpha) = -\partial_\beta \Pi_{\alpha\beta}^{(1)} - \partial_{t_0}\sum_i c_{i\alpha} f_i^{(1)} - \frac{1}{2}\partial_{t_0 t_0}(\rho u_\alpha) - \frac{1}{2}\partial_{\beta\gamma}P_{\alpha\beta\gamma}^{(0)} - \partial_{t_0\beta}\Pi_{\alpha\beta}^{(0)} \quad (A25)$$

$$\partial_{t_1}\rho(bT + u_\alpha^2) = -\partial_\beta Q_\beta^{(1)} - \partial_{t_0}\sum_i (c_{i\alpha}^2 + \eta_i^2) f_i^{(1)}$$
$$-\frac{1}{2}\partial_{t_0 t_0}\rho(bT + u_\alpha^2) - \frac{1}{2}\partial_{\beta\gamma}R_{\beta\gamma}^{(0)} - \partial_{t_0\beta}Q_\beta^{(0)} \quad (A26)$$

Recover the time derivative, then

$$\partial_t \rho + \partial_\beta (\rho u_\beta) = 0 \quad (A27)$$

$$\partial_t (\rho u_\alpha) + \partial_\beta \Pi_{\alpha\beta} = \delta\left(\frac{1}{\omega} - \frac{1}{2}\right)\left(\partial_{t_0\beta}\Pi_{\alpha\beta}^{(0)} + \partial_{\beta\gamma}P_{\alpha\beta\gamma}^{(0)}\right) \quad (A28)$$

$$\partial_t \rho(bT + u_\alpha^2) + \partial_\beta Q_\beta^{(0)} = \delta\left(\frac{1}{\omega} - \frac{1}{2}\right)\left(\partial_{t_0\beta}Q_\beta^{(0)} + \partial_{\beta\gamma}R_{\beta\gamma}^{(0)}\right) \quad (A29)$$

Now we study on the one dimensional model. Based on Eq.(A19)-(A21), The Euler equations are

$$\frac{\partial \rho}{\partial t_0} + \frac{\partial (\rho u)}{\partial x} = 0 \quad (A30)$$

$$\frac{\partial (\rho u)}{\partial t_0} + \frac{\partial (\rho u^2 + p)}{\partial x} = 0 \quad (A31)$$

$$\frac{\partial \rho(bT + u^2)}{\partial t_0} + \frac{\partial \left[\rho u(bT + u^2) + 2pu\right]}{\partial x} = 0 \quad (A32)$$

Where

$$p = \rho T \quad (A33)$$

Is the equation of state for ideal gas.

Based on Eqs.(A27)-(A29) and Eqs.(18c)-(18e), The Euler equations become

$$\frac{\partial \rho}{\partial t} + \frac{\partial (\rho u)}{\partial x} = 0 \tag{A34}$$

$$\frac{\partial (\rho u)}{\partial t} + \frac{\partial (\rho u^2 + p)}{\partial x} = R_2 \tag{A35}$$

$$\frac{\partial (\rho u^2 + \rho bT)}{\partial t} + \frac{\partial \rho[(b+2)T + u^2]u}{\partial x} = R_3 \tag{A36}$$

Where

$$R_2 = \delta\left(\frac{1}{\omega} - \frac{1}{2}\right)\left[\frac{\partial^2 (\rho u^2 + p)}{\partial t_0 \partial x} + \frac{\partial^2 \left(\sum c_i c_i c_i f_i^{eq}\right)}{\partial x \partial x}\right] \tag{A37}$$

$$R_3 = \delta\left(\frac{1}{\omega} - \frac{1}{2}\right)\left[\frac{\partial^2 \rho[(b+2)T + u^2]u}{\partial t_0 \partial x} + \frac{\partial^2 \left[\sum (c_i^2 + \eta_i^2) c_i c_i f_i^{eq}\right]}{\partial x \partial x}\right] \tag{A38}$$

Now we compute $R_2$ and $R_3$ as follows

Based on Eq.(19)-Eq.(21b)

$$\sum c_i c_i c_i f_i^{eq}$$
$$= v_1^3 \left(f_1^{eq} - f_2^{eq}\right) + v_2^3 \left(f_3^{eq} - f_4^{eq}\right)$$
$$= v_1^3 \rho\left[(A_1 + B_1 u_1 v_1) - (A_2 - B_2 u_1 v_1)\right] + v_2^3 \rho\left[(A_3 + B_3 u_1 v_2) - (A_4 - B_4 u_1 v_2)\right]$$
$$= v_1^3 \rho 2 B_1 u_1 v_1 + v_2^3 \rho 2 B_3 u_1 v_2$$
$$= 2\rho u_1 \left(B_1 v_1^4 + B_3 v_2^4\right)$$
$$= (b+2) pu + \rho u^3 \tag{A39}$$

$$\sum (c_i^2 + \eta_i^2) c_i c_i f_i^{eq}$$
$$= v_1^4 \left(f_1^{eq} + f_2^{eq}\right) + v_2^4 \left(f_3^{eq} + f_4^{eq}\right)$$
$$= v_1^4 \rho\left[(A_1 + B_1 u_1 v_1) + (A_2 - B_2 u_1 v_1)\right] + v_2^4 \rho\left[(A_3 + B_3 u_1 v_2) + (A_4 - B_4 u_1 v_2)\right]$$

$$= v_1^4 \rho 2A_1 + v_2^4 \rho 2A_3$$

$$= 2\rho\left(A_1 v_1^4 + A_3 v_2^4\right)$$

$$= -v_1^2 v_2^2 \rho + \frac{(b-1)v_1^2 v_2^2}{\eta_0^2}p + \left(v_1^2 + v_2^2\right)\left(\rho u^2 + p\right) \tag{A40}$$

Based on Eq.(A30)-A(32)

$$\frac{\partial\left(\rho u^2 + p\right)}{\partial t_0}$$

$$= u\frac{\partial \rho u}{\partial t_0} + \rho u\frac{\partial u}{\partial t_0} + \frac{1}{b}\left[-\frac{\partial\left(\rho u\left(bT + u^2\right) + 2pu\right)}{\partial x} - \frac{\partial \rho u^2}{\partial t_0}\right]$$

$$= u\frac{\partial \rho u}{\partial t_0} + u\left(\frac{\partial \rho u}{\partial t_0} - u\frac{\partial \rho}{\partial t_0}\right) + \frac{1}{b}\left[-\frac{\partial\left(\rho u\left(bT + u^2\right) + 2pu\right)}{\partial x} + 2u\frac{\partial\left(\rho u^2 + p\right)}{\partial x} - u^2\frac{\partial \rho u}{\partial x}\right]$$

$$= 2u\frac{\partial \rho u}{\partial t_0} - u^2\frac{\partial \rho}{\partial t_0} + \frac{1}{b}\left[-\frac{\partial\left(\rho u\left(bT + u^2\right) + 2pu\right)}{\partial x} + 2u\frac{\partial\left(\rho u^2 + p\right)}{\partial x} - u^2\frac{\partial \rho u}{\partial x}\right]$$

$$= -2u\frac{\partial\left(\rho u^2 + p\right)}{\partial x} + u^2\frac{\partial \rho u}{\partial x}$$

$$+ \frac{1}{b}\left[-\frac{\partial\left(\rho u\left(bT + u^2\right) + 2pu\right)}{\partial x} + 2u\frac{\partial\left(\rho u^2 + p\right)}{\partial x} - u^2\frac{\partial \rho u}{\partial x}\right] \tag{A41}$$

$$\frac{\partial \rho\left[(b+2)T + u^2\right]u}{\partial t_0}$$

$$= (b+2)\frac{\partial pu}{\partial t_0} + \frac{\partial \rho u^3}{\partial t_0}$$

$$= (b+2)\left(u\frac{\partial p}{\partial t_0} + p\frac{\partial u}{\partial t_0}\right) + u\frac{\partial \rho u^2}{\partial t_0} + \rho u^2\frac{\partial u}{\partial t_0}$$

$$= (b+2)\left[u\frac{\partial p}{\partial t_0} + \frac{p}{\rho}\left(\frac{\partial \rho u}{\partial t_0} - u\frac{\partial \rho}{\partial t_0}\right)\right] + u\frac{\partial \rho u^2}{\partial t_0} + u^2\left(\frac{\partial \rho u}{\partial t_0} - u\frac{\partial \rho}{\partial t_0}\right)$$

$$= \frac{u(b+2)}{b}\left[-\frac{\partial\left(\rho u\left(bT+u^2\right)+2pu\right)}{\partial x}+2u\frac{\partial\left(\rho u^2+p\right)}{\partial x}-u^2\frac{\partial \rho u}{\partial x}\right]+\frac{p(b+2)}{\rho}\left(-\frac{\partial\left(\rho u^2+p\right)}{\partial x}+u\frac{\partial \rho u}{\partial x}\right)$$

$$+u\left(-2u\frac{\partial\left(\rho u^2+p\right)}{\partial x}+u^2\frac{\partial \rho u}{\partial x}\right)+u^2\left[-\frac{\partial\left(\rho u^2+p\right)}{\partial x}+u\frac{\partial \rho u}{\partial x}\right]$$

$$= \frac{u(b+2)}{b}\left[-\frac{\partial\left(\rho u\left(bT+u^2\right)+2pu\right)}{\partial x}+2u\frac{\partial\left(\rho u^2+p\right)}{\partial x}-u^2\frac{\partial \rho u}{\partial x}\right]+\frac{p(b+2)}{\rho}\left(-\frac{\partial\left(\rho u^2+p\right)}{\partial x}+u\frac{\partial \rho u}{\partial x}\right)$$

$$-3u^2\frac{\partial\left(\rho u^2+p\right)}{\partial x}+2u^3\frac{\partial \rho u}{\partial x} \tag{A42}$$

Based on Eq.(A39)-(A42)

$$R_2 = \delta\left(\frac{1}{\omega}-\frac{1}{2}\right)\left[\frac{\partial^2\left(\rho u^2+p\right)}{\partial t_0 \partial x}+\frac{\partial^2\left(\sum c_i c_i c_i f_i^{eq}\right)}{\partial x \partial x}\right]$$

$$= \delta\left(\frac{1}{\omega}-\frac{1}{2}\right)\left\{2\left(\frac{1}{b}-1\right)\left[\frac{\partial u}{\partial x}\frac{\partial\left(\rho u^2+p\right)}{\partial x}+u\frac{\partial^2\left(\rho u^2+p\right)}{\partial x^2}\right]+\left(1-\frac{1}{b}\right)\left[2u\frac{\partial u}{\partial x}\frac{\partial \rho u}{\partial x}+u^2\frac{\partial^2 \rho u}{\partial x^2}\right]\right.$$
$$\left.-\frac{1}{b}\frac{\partial^2\left[\rho u\left(bT+u^2\right)+2pu\right]}{\partial x^2}+\partial_{xx}\left[2\rho u\left(B_1 v_1^4+B_3 v_2^4\right)\right]\right\}$$

$$= \delta\left(\frac{1}{\omega}-\frac{1}{2}\right)\left\{\begin{array}{l}2\left(\dfrac{1}{b}-1\right)\left(4\rho u\dfrac{\partial u}{\partial x}\dfrac{\partial u}{\partial x}+5u^2\dfrac{\partial \rho}{\partial x}\dfrac{\partial u}{\partial x}+\dfrac{\partial p}{\partial x}\dfrac{\partial u}{\partial x}+2\rho u^2\dfrac{\partial^2 u}{\partial x^2}+u^3\dfrac{\partial^2 \rho}{\partial x^2}+u\dfrac{\partial^2 p}{\partial x^2}\right)\\ +\left(1-\dfrac{1}{b}\right)\left(2\rho u\dfrac{\partial u}{\partial x}\dfrac{\partial u}{\partial x}+4u^2\dfrac{\partial \rho}{\partial x}\dfrac{\partial u}{\partial x}+\rho u^2\dfrac{\partial^2 u}{\partial x^2}+u^3\dfrac{\partial^2 \rho}{\partial x^2}\right)\\ +\left(1-\dfrac{1}{b}\right)\left[\begin{array}{l}(b+2)\left(p\dfrac{\partial^2 u}{\partial x^2}+2\dfrac{\partial p}{\partial x}\dfrac{\partial u}{\partial x}+u\dfrac{\partial^2 p}{\partial x^2}\right)+u^3\dfrac{\partial^2 \rho}{\partial x^2}+6u^2\dfrac{\partial \rho}{\partial x}\dfrac{\partial u}{\partial x}\\ +6\rho u\dfrac{\partial u}{\partial x}\dfrac{\partial u}{\partial x}+3\rho u^2\dfrac{\partial^2 u}{\partial x^2}\end{array}\right]\end{array}\right\}$$

$$= \delta\left(\frac{1}{\omega}-\frac{1}{2}\right)\left\{\begin{array}{l}2\left(\dfrac{1}{b}-1\right)\left(4\rho u\dfrac{\partial u}{\partial x}\dfrac{\partial u}{\partial x}+5u^2\dfrac{\partial \rho}{\partial x}\dfrac{\partial u}{\partial x}+\dfrac{\partial p}{\partial x}\dfrac{\partial u}{\partial x}+2\rho u^2\dfrac{\partial^2 u}{\partial x^2}+u^3\dfrac{\partial^2 \rho}{\partial x^2}+u\dfrac{\partial^2 p}{\partial x^2}\right)\\ -\left(\dfrac{1}{b}-1\right)\left[\begin{array}{l}(b+2)\left(p\dfrac{\partial^2 u}{\partial x^2}+2\dfrac{\partial p}{\partial x}\dfrac{\partial u}{\partial x}+u\dfrac{\partial^2 p}{\partial x^2}\right)+2u^3\dfrac{\partial^2 \rho}{\partial x^2}+10u^2\dfrac{\partial \rho}{\partial x}\dfrac{\partial u}{\partial x}\\ +8\rho u\dfrac{\partial u}{\partial x}\dfrac{\partial u}{\partial x}+4\rho u^2\dfrac{\partial^2 u}{\partial x^2}\end{array}\right]\end{array}\right\}$$

$$= \delta\left(\frac{1}{\omega}-\frac{1}{2}\right)\left\{\left(\frac{1}{b}-1\right)\left(2\frac{\partial p}{\partial x}\frac{\partial u}{\partial x}+2u\frac{\partial^2 p}{\partial x^2}\right)-\left(\frac{1}{b}-1\right)\left[(b+2)\left(p\frac{\partial^2 u}{\partial x^2}+2\frac{\partial p}{\partial x}\frac{\partial u}{\partial x}+u\frac{\partial^2 p}{\partial x^2}\right)\right]\right\}$$

$$= \delta\left(\frac{1}{\omega}-\frac{1}{2}\right)\left[\frac{2(b^2-1)}{b}\frac{\partial p}{\partial x}\frac{\partial u}{\partial x}+(b-1)u\frac{\partial^2 p}{\partial x^2}+\frac{(b+2)(b-1)}{b}p\frac{\partial^2 u}{\partial x^2}\right] \quad (A43)$$

$$R_3 = \delta\left(\frac{1}{\omega}-\frac{1}{2}\right)\left[\frac{\partial^2 \rho\left[(b+2)T+u^2\right]u}{\partial t_0 \partial x}+\frac{\partial^2\left[\sum(c_i^2+\eta_i^2)c_ic_if_i^{eq}\right]}{\partial x \partial x}\right]$$

$$= \delta\left(\frac{1}{\omega}-\frac{1}{2}\right)\left\{\begin{array}{l} -\dfrac{(b+2)}{b}\left[\dfrac{\partial u}{\partial x}\dfrac{\partial(bpu+\rho u^3+2pu)}{\partial x}+u\dfrac{\partial^2(bpu+\rho u^3+2pu)}{\partial x^2}\right] \\[6pt] +\left[\dfrac{4(b+2)u}{b}\dfrac{\partial u}{\partial x}-\dfrac{(b+2)}{\rho^2}\left(\rho\dfrac{\partial p}{\partial x}-p\dfrac{\partial \rho}{\partial x}\right)-6u\dfrac{\partial u}{\partial x}\right]\dfrac{\partial(\rho u^2+p)}{\partial x} \\[6pt] +\left(\dfrac{2(b+2)u^2}{b}-\dfrac{(b+2)p}{\rho}-3u^2\right)\dfrac{\partial^2(\rho u^2+p)}{\partial x^2} \\[6pt] +\left[6u^2\dfrac{\partial u}{\partial x}+\dfrac{(b+2)}{\rho^2}\left(\rho\dfrac{\partial pu}{\partial x}-pu\dfrac{\partial \rho}{\partial x}\right)-\dfrac{3(b+2)u^2}{b}\dfrac{\partial u}{\partial x}\right]\dfrac{\partial \rho u}{\partial x} \\[6pt] +\left(2u^3+\dfrac{(b+2)pu}{\rho}-\dfrac{(b+2)u^3}{b}\right)\dfrac{\partial^2 \rho u}{\partial x^2}+\partial_{xx}\left[2\rho(A_1 v_1^4+A_3 v_2^4)\right] \end{array}\right\}$$

$$= \delta\left(\frac{1}{\omega}-\frac{1}{2}\right)\left\{\begin{array}{l} -\dfrac{(b+2)}{b}\left[\dfrac{\partial u}{\partial x}\dfrac{\partial(bpu+\rho u^3+2pu)}{\partial x}+u\dfrac{\partial^2(bpu+\rho u^3+2pu)}{\partial x^2}\right] \\[6pt] +\left[\dfrac{4(b+2)u}{b}\dfrac{\partial u}{\partial x}-\dfrac{(b+2)}{\rho^2}\left(\rho\dfrac{\partial p}{\partial x}-p\dfrac{\partial \rho}{\partial x}\right)-6u\dfrac{\partial u}{\partial x}\right]\dfrac{\partial(\rho u^2+p)}{\partial x} \\[6pt] +\left(\dfrac{2(b+2)u^2}{b}-\dfrac{(b+2)p}{\rho}-3u^2\right)\dfrac{\partial^2(\rho u^2+p)}{\partial x^2} \\[6pt] +\left[6u^2\dfrac{\partial u}{\partial x}+\dfrac{(b+2)}{\rho^2}\left(\rho\dfrac{\partial pu}{\partial x}-pu\dfrac{\partial \rho}{\partial x}\right)-\dfrac{3(b+2)u^2}{b}\dfrac{\partial u}{\partial x}\right]\dfrac{\partial \rho u}{\partial x} \\[6pt] +\left(2u^3+\dfrac{(b+2)pu}{\rho}-\dfrac{(b+2)u^3}{b}\right)\dfrac{\partial^2 \rho u}{\partial x^2} \\[6pt] +\dfrac{\partial^2\left[-v_1^2 v_2^2 \rho+\dfrac{(b-1)v_1^2 v_2^2}{\eta_0^2}p+(v_1^2+v_2^2)(\rho u^2+p)\right]}{\partial x^2} \end{array}\right\}$$

$$= \delta\left(\frac{1}{\omega} - \frac{1}{2}\right) \begin{Bmatrix} -\left[\frac{2(b+1)(b+2)}{b}p + 12\rho u^2 + 2\rho(v_1^2 + v_2^2)\right]\left(\frac{\partial u}{\partial x}\right)^2 \\ -\frac{4(b^2 + 4b + 1)}{b}u\frac{\partial p}{\partial x}\frac{\partial u}{\partial x} - \left[8u^3 - 4u(v_1^2 + v_2^2)\right]\frac{\partial \rho}{\partial x}\frac{\partial u}{\partial x} \\ +(b+2)\frac{p}{\rho^2}\frac{\partial \rho}{\partial x}\frac{\partial p}{\partial x} - (b+2)\frac{1}{\rho}\frac{\partial p}{\partial x}\frac{\partial p}{\partial x} \\ -\left[\frac{2(b+2)(b+1)}{b}pu + 4\rho u^3 + 2\rho u(v_1^2 + v_2^2)\right]\frac{\partial^2 u}{\partial x^2} \\ -\left[(b+5)u^2 + (b+2)\frac{p}{\rho} + \frac{(b-1)v_1^2 v_2^2}{\eta_0^2} + (v_1^2 + v_2^2)\right]\frac{\partial^2 p}{\partial x^2} \\ -\left[u^4 + v_1^2 v_2^2 + (v_1^2 + v_2^2)u^2\right]\frac{\partial^2 \rho}{\partial x^2} \end{Bmatrix} \quad (A44)$$

$$R_3 - uR_2 = \delta\left(\frac{1}{\omega} - \frac{1}{2}\right) \begin{Bmatrix} -\left[\frac{2(b+1)(b+2)}{b}p + 12\rho u^2 + 2\rho(v_1^2 + v_2^2)\right]\left(\frac{\partial u}{\partial x}\right)^2 \\ -\frac{2(3b^2 + 8b + 1)}{b}u\frac{\partial p}{\partial x}\frac{\partial u}{\partial x} - \left[8u^3 - 4u(v_1^2 + v_2^2)\right]\frac{\partial \rho}{\partial x}\frac{\partial u}{\partial x} \\ +(b+2)\frac{p}{\rho^2}\frac{\partial \rho}{\partial x}\frac{\partial p}{\partial x} - (b+2)\frac{1}{\rho}\frac{\partial p}{\partial x}\frac{\partial p}{\partial x} \\ -\left[\frac{(b+2)(3b+1)}{b}pu + 4\rho u^3 + 2\rho u(v_1^2 + v_2^2)\right]\frac{\partial^2 u}{\partial x^2} \\ -\left[2(b+2)u^2 + (b+2)\frac{p}{\rho} + \frac{(b-1)v_1^2 v_2^2}{\eta_0^2} + (v_1^2 + v_2^2)\right]\frac{\partial^2 p}{\partial x^2} \\ -\left[u^4 + v_1^2 v_2^2 + (v_1^2 + v_2^2)u^2\right]\frac{\partial^2 \rho}{\partial x^2} \end{Bmatrix} \quad (A45)$$

If we use finite difference method to compute kinetic equation Eqs.(15)-(16),Eq.(A45) become

$$R_3 - uR_2 = \frac{\delta}{\omega} \begin{Bmatrix} -\left[\frac{2(b+1)(b+2)}{b}p + 12\rho u^2 + 2\rho\left(v_1^2 + v_2^2\right)\right]\left(\frac{\partial u}{\partial x}\right)^2 \\ -\frac{2(3b^2 + 8b + 1)}{b}u\frac{\partial p}{\partial x}\frac{\partial u}{\partial x} - \left[8u^3 - 4u\left(v_1^2 + v_2^2\right)\right]\frac{\partial \rho}{\partial x}\frac{\partial u}{\partial x} \\ +(b+2)\frac{p}{\rho^2}\frac{\partial \rho}{\partial x}\frac{\partial p}{\partial x} - (b+2)\frac{1}{\rho}\frac{\partial p}{\partial x}\frac{\partial p}{\partial x} \\ -\left[\frac{(b+2)(3b+1)}{b}pu + 4\rho u^3 + 2\rho u\left(v_1^2 + v_2^2\right)\right]\frac{\partial^2 u}{\partial x^2} \\ -\left[2(b+2)u^2 + (b+2)\frac{p}{\rho} + \frac{(b-1)v_1^2 v_2^2}{\eta_0^2} + \left(v_1^2 + v_2^2\right)\right]\frac{\partial^2 p}{\partial x^2} \\ -\left[u^4 + v_1^2 v_2^2 + \left(v_1^2 + v_2^2\right)u^2\right]\frac{\partial^2 \rho}{\partial x^2} \end{Bmatrix} \quad (A46)$$